\documentclass[pra,twocolumn,showpacs,
	floatfix, reprint]{revtex4-2}
	\usepackage{subcaption}
	\usepackage[usenames,dvipsnames]{color}
	\usepackage{float}
	\usepackage{graphicx}
	\usepackage{color}
	\usepackage{booktabs}
	\usepackage{amsmath}
	\usepackage{multirow}
	\usepackage{amssymb,amsmath,verbatim}
	\usepackage[normalem]{ulem}
	\usepackage{booktabs}   
	
	\usepackage{ragged2e}

\newcommand{\emdash}{\unskip---\ignorespaces}
	\newcommand{\ket}[1]{|{#1}\rangle}
	\newcommand{\bra}[1]{\langle{#1}|}

	\usepackage{hyperref}
	\hypersetup{
	  colorlinks   = true, 
	  urlcolor     = blue, 
	  linkcolor    = blue, 
	  citecolor   = blue 
	}
	
\begin{document}
	
	

    \title{Efficient direct loading of the green MOT of Yb with low green laser power}
	
	
	 \author{Thilagaraj Ravi}
  \author{Rajnandan Choudhury Das}
	 \author{Heramb Vivek Bhusane}
	 \author{Samrat Roy}
	\author{Kanhaiya Pandey}%
	\email{kanhaiyapandey@iitg.ac.in}
	\affiliation{%
	 Department of Physics, Indian Institute of Technology Guwahati, Guwahati, Assam 781039, India
	}%
	
	\date{\today}

\begin{abstract}
We report the direct loading of Ytterbium (Yb) atoms in magneto-optical trap (MOT) using the intercombination narrow optical transition 6s$^2$ $^1$S$_0$ $\rightarrow$ 6s6p $^3$P$_1$ at 556 nm (green MOT) with limited green laser power of  10 mW. Direct loading of the green MOT is achieved by superimposing the green laser beam, inside a hollow core of the laser beam driving the broad 6s$^2$ $^1$S$_0$ $\rightarrow$ 6s6p $^1$P$_1$ transition (blue) at 399 nm. We load up to 3$\times10^8$ atoms in $1$ s. We characterize the green MOT loading with various experimental parameters such as magnetic field gradient, power of the green laser and blue MOT laser, and detuning of the green laser. We have also loaded the green MOT using center-shifted dual MOT configuration. In this configuration, the overlap region of the three counter-propagating blue laser beams is shifted towards Zeeman slower, where the magnetic field is non-zero. The atoms are first pre-cooled and partially trapped in the blue MOT. These atoms enter the green MOT region and are trapped. In this method, unlike in the core-shell MOT, we do not lose blue MOT laser power due to masking of the central region. However, we load only 10$^7$ atoms, which is one order of magnitude lower than that in the core-shell MOT.    

\end{abstract}

\maketitle
\section{Introduction}

Alkaline-earth elements, such as Strontium (Sr) and Alkaline-earth-like elements, such as Ytterbium (Yb), have attracted significant interest within the field of cold atom physics due to their complex electronic structures and versatile transition properties because of their double-valence electrons. Unlike single-valence-electron species, two-valence-electron atoms offer a spectrum of optical transitions\emdash ranging from broad to ultra narrow line widths \emdash which enhance their utility across a variety of cutting-edge applications such as quantum precision measurements \cite{Poli2011}, serve as references in optical atomic clocks \cite{Ludlow2013, Ivaylo2019}, and play pivotal roles in quantum simulations \cite{covey2018,kaufman_qs} and computation \cite{Manuel2018,kaufman2022,covey2023,Thompson_qc}. The absence of electron spin in certain isotopes, or the existence of purely nuclear hyperfine levels, further extends their suitability for applications where minimal decoherence is desired. These elements have also emerged as an ideal candidate for exploring Rydberg states for quantum computation and simulation, where the dual valence electrons facilitate simultaneous trapping of ground and excited states in optical tweezers \cite{Thompson2022}, as well as optical imaging of Rydberg atoms \cite{ThompsonPRX}. 

  For the above applications, one of the important steps is the cooling and trapping in magneto-optical trap (MOT) using narrow inter-combination line, as it provides low temperature. However, trapping at this transition is challenging because of low capture velocity. Two-step MOT is used for loading the atoms at narrow transition, where a broad transition is first employed to rapidly capture atoms up to high velocity. This is followed by transfer to the MOT at narrow transition by switching off the broad transition laser beam and lowering the magnetic field gradient. The MOT using narrow transition provides lower temperatures and higher density clouds \cite{Yabuzaki1999, Natarajan2010, Wilkowski2015, Benjamin2011, Pfau2014, Thywissen2011, hulet2011, rajnandan2023, Mikio2021}. In many cases, such as in Yb and Sr, the broad transition is not completely close and limits the number of atoms in the MOT. Therefore, it is preferable to directly load atoms into the MOT using a narrow-linewidth transition, as this transition happens to be close.

In the case of Yb, direct loading has been demonstrated using the green inter-combination transition at 556 nm (green MOT). The green MOT loads a larger number of atoms, but with a slow loading rate \cite{Guttridge_2016}. The direct loading of the green MOT is possible by utilizing the power broadening effect or broadening the laser linewidth using an acousto-optic modulator (AOM) or an electro-optic modulator (EOM), but requires high power of the green laser \cite{TakahashiGreenMot}. The green laser at 556 nm is not directly available from diode lasers and is typically generated either by frequency doubling of an infrared laser or using a dye laser pumped by an argon-ion laser \cite{TakahashiGreenMot}. To generate high green power (on the order of a few hundred mW), frequency doubling is performed using a cavity-enhanced setup, where a nonlinear crystal is placed inside a resonant cavity. This cavity increases the circulating power of the fundamental laser, thereby improving the conversion to green light. However, lower green powers (a few tens of mW) can be achieved using a simpler single-pass frequency doubling through a nonlinear crystal \cite{Yamaguchi2008}. For applications such as portable atomic clocks, it is desired to avoid the frequency-doubling cavity and hence to load the green MOT even with low power of the green laser. Core-shell configuration \cite{Mun2015, RCD_CMOT}, where a hollow core is created (by masking) in the laser beam of the broad transition and filled by the weak transition laser beam, has been very useful for loading the green MOT faster.

In this work, we load the green MOT of Yb using low green laser power in the core-shell configuration. We create a hollow core in the blue laser beam such that the masked (using a mirror) portion can be utilized for 2D cooling. We also explore the other configuration of center-shifted dual MOT. In this configuration, the overlap region of the three counter-propagating broad transition (blue) laser beams is shifted towards Zeeman slower with non-zero magnetic field. The atoms are pre-cooled and partially trapped by the blue transition as the magnetic field is non-zero. These atoms are trapped in the green MOT once they enter the green MOT region.

The paper is organized as follows. In section \ref{ExpSetUp}, we describe the experimental set-up, which includes our vacuum setup, laser system, and optics scheme. In section \ref{Theory}, we discuss about the theoretical model to study the capture dynamics of atoms. In section \ref{RnD}, we describe our experimental results. In section \ref{Conclusion}, we summarize our findings.

\section{Experimental Set- up}\label{ExpSetUp}

\subsection{Vacuum setup}
\begin{figure*}[]
	 \includegraphics[width=0.9\linewidth]{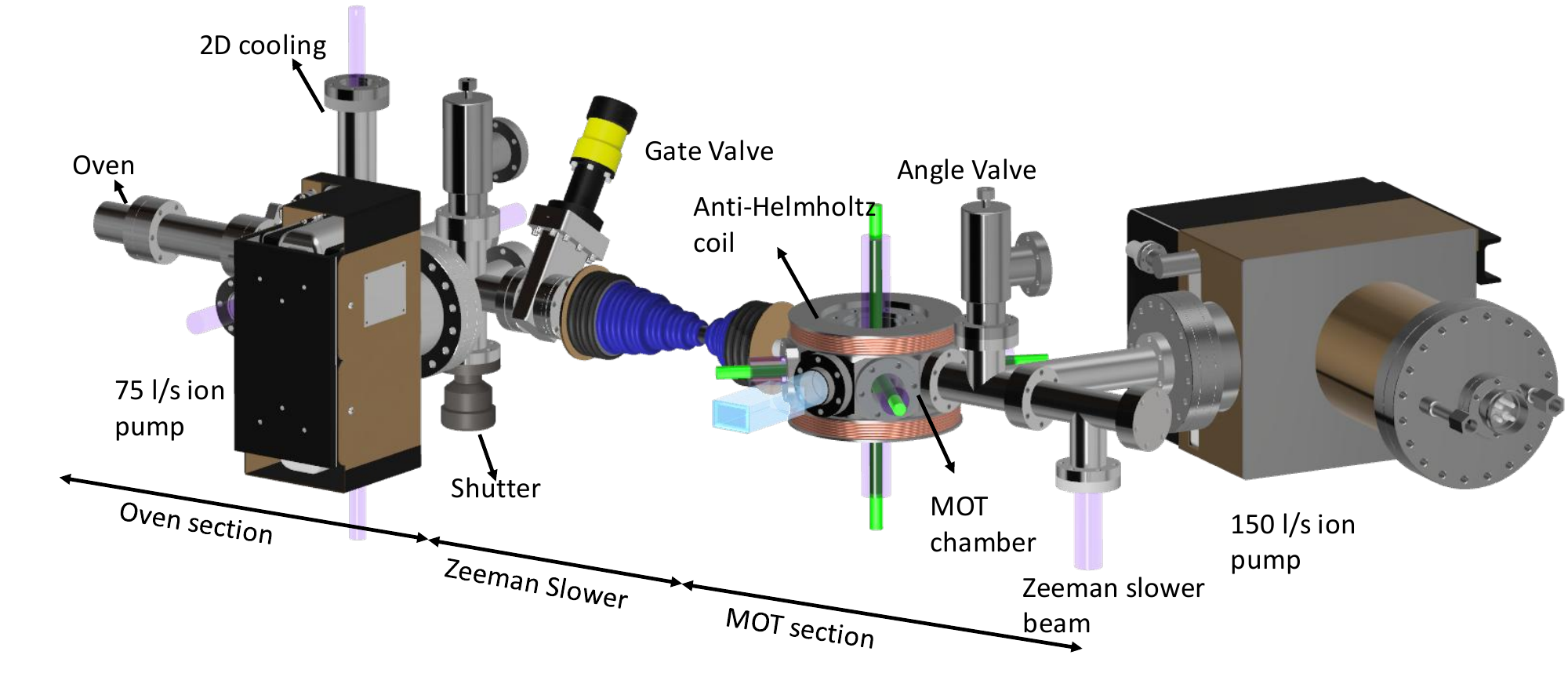}
    \includegraphics[width=\linewidth]{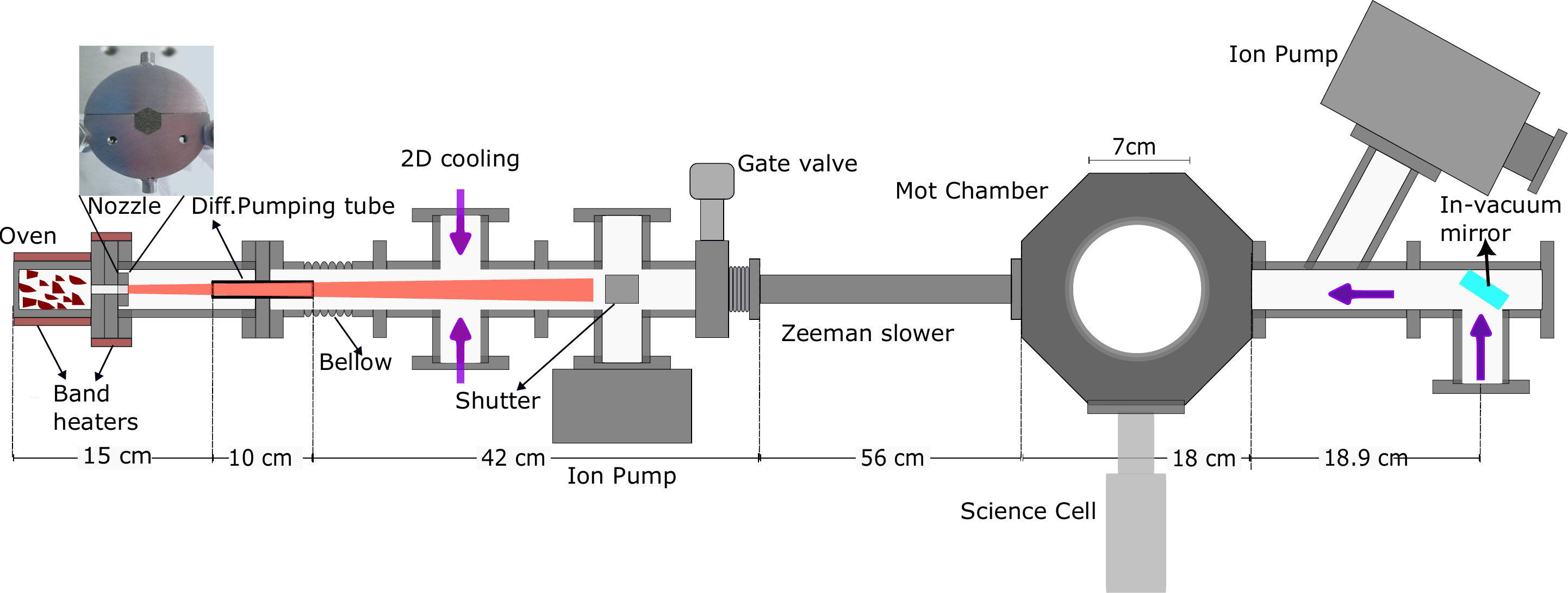}
   	\begin{picture}(0,0)
		\put(-220,380){(a)}
		\put(-220,202){(b)}
	\end{picture}
	\caption{\label{Vs} \justifying (a) CAD design of our vacuum system. It comprises various sections: oven, 2D cooling, Zeeman slower, and MOT chamber. (b) 2-dimensional view of the vacuum system. }
\end{figure*}
 The overview of our experimental vacuum system is shown in Fig. \ref{Vs}.
\subsubsection{Yb oven}
Yb atomic beam is produced from the oven operating at 400 $^{o}$C. The oven is made of a stainless steel crucible attached with a CF40 bored flange of bore diameter 10 mm. It is filled with 25 g of Yb chunks (Thermo Scientific Chemicals). The Yb atoms vaporized from the oven are collimated by the micro-capillaries. Each capillary tube (Coopers Needle Works Ltd) is made of stainless steel and
has a length of 10 mm with inner and outer diameters of 190 $\mu$m and
300 $\mu$m, respectively. The micro-capillaries are stacked into a hexagonal aperture of 3 mm side length, machined at the center of a cylindrical disc assembly as shown in Fig. \ref{Vs}(b). The disc is composed of two stainless steel parts held together with screws. The purpose of this micro-capillary nozzle is to increase the lifespan of the oven and facilitate the well-collimation of the atomic beam. The oven crucible and nozzle are heated using ceramic band heaters. The temperature of the nozzle is always maintained at 50 $^{o}$C higher than the crucible to avoid the clogging of capillaries. The atomic beam is further collimated by a differential pumping tube located at 66 mm from the micro capillary nozzle as shown in Fig. \ref{Vs}(b). The inner diameter and the length of the differential pumping tube are 8 mm and 100 mm, respectively.  The oven is attached to the rest of the vacuum system through a CF40 bellow of length 90 mm to adjust the direction of the atomic beam towards the Zeeman slower and 3D MOT.

\subsubsection{2D cooling}
After the oven, there is a 2-dimensional (2D) transverse cooling stage.   
The 2D cooling stage is a four-way cross with arm length of 185 mm. It has four windows AR-coated for 399 nm (VPZ38VAR, Torr Scientific). The extended arm length prevents the windows from the deposition of Yb atoms. Another four-way cross next to the 2D cooling section features the ion pump (Vacion Plus 75 starcell, Agilent), angle valve and motorized rotary feed-through (BRM-275-03, MDC Precision). The rotary feed-through houses an atomic beam shutter. The shutter is a stainless steel hollow square block with two opposite faces open, allowing the atomic beam to pass through when aligned. It blocks and transmits the atomic beam every {90}$^o$ rotation. The oven and 2D cooling section can be isolated from the remaining part of the vacuum chamber using a gate valve (E-GV-1500M-11, MDC Precision ). This will be useful for the refilling of the oven. The pressure in the oven section is around $2\times10^{-8}$ mbar during the oven operation. 

During the first round of baking, the 2D cooling stage windows became opaque due to the coating of Yb atoms. This is because the 2D cooling portion was baked when the oven was at a high temperature (380 $^o$C), which caused the atoms to diffuse toward and coat the windows. To avoid this issue, in the second attempt, the 2D cooling section was kept at room temperature while the oven was at a high temperature. Later, when the temperature of the 2D cooling portion was increased to 150 $^o$C, the oven temperature was lowered to 270 $^o$C.          

\subsubsection{Zeeman slower}

  The collimated atomic beam is slowed down by the spin-flip Zeeman slower, which comprises a CF16 nipple of length 55 cm. The CF16 nipple helps in further collimation of the atomic beam. To generate the Zeeman slowing magnetic field profile, two types of wires are wound around a 50 cm long CF40 nipple, concentric with the CF16 slower nipple. A high-current wire (diameter 11 mm), capable of carrying up to 130 A, is wound at both ends of the CF40 nipple. A lower-current wire (diameter 7 mm), carrying up to 30 A, is wound around the central portion of the CF40 nipple, as shown in Fig.~\ref{Vs}(a). Due to the gap between the CF16 and CF40 nipples, heat from the current-carrying coils is dissipated without much transfer to the vacuum. The magnetic field is generated at the beginning and at the end of the slower nipple are $+$265 Gauss and $-$360 Gauss, respectively. With a $-$500 MHz detuning of the slower laser, the Zeeman slower can slow the atom with velocities up to 350 m/s. Instead of sending the slowing laser beam through a window directly facing the Yb atomic beam, we use an in-vacuum mirror (PFE10-P01, Thorlabs) positioned at an angle of 45$^{o}$. This arrangement helps us to avoid Yb deposition on the window.    
\subsubsection{3D MOT}
An octagon chamber is connected next to the Zeeman slower nipple, where the Yb atoms are cooled and trapped using MOT beams in retro configuration. The dimensions of the chamber are 70 mm x 70 mm (side length $\times$ height). It has two CF63 viewports (Make: Torr Scientific, Model: VPZ64BBAR-LN) and eight CF40 viewports (Make: Torr Scientific, Model: VPZ38BBAR-LN) with clear apertures of 63 mm and 38 mm, respectively. A pair of copper coils generates the magnetic field gradient 8~G/cm per ampere in anti-Helmholtz configuration. Each coil has an inner diameter of 124 mm, an outer diameter of 200 mm, and 650 turns. The MOT chamber is maintained at a pressure around  $5\times10^{-10}$ mbar during operation, pumped by a 150 l/s ion pump equipped with a Ti-sublimator.  

 \begin{figure}[htb]
 	  \includegraphics[width=\linewidth]{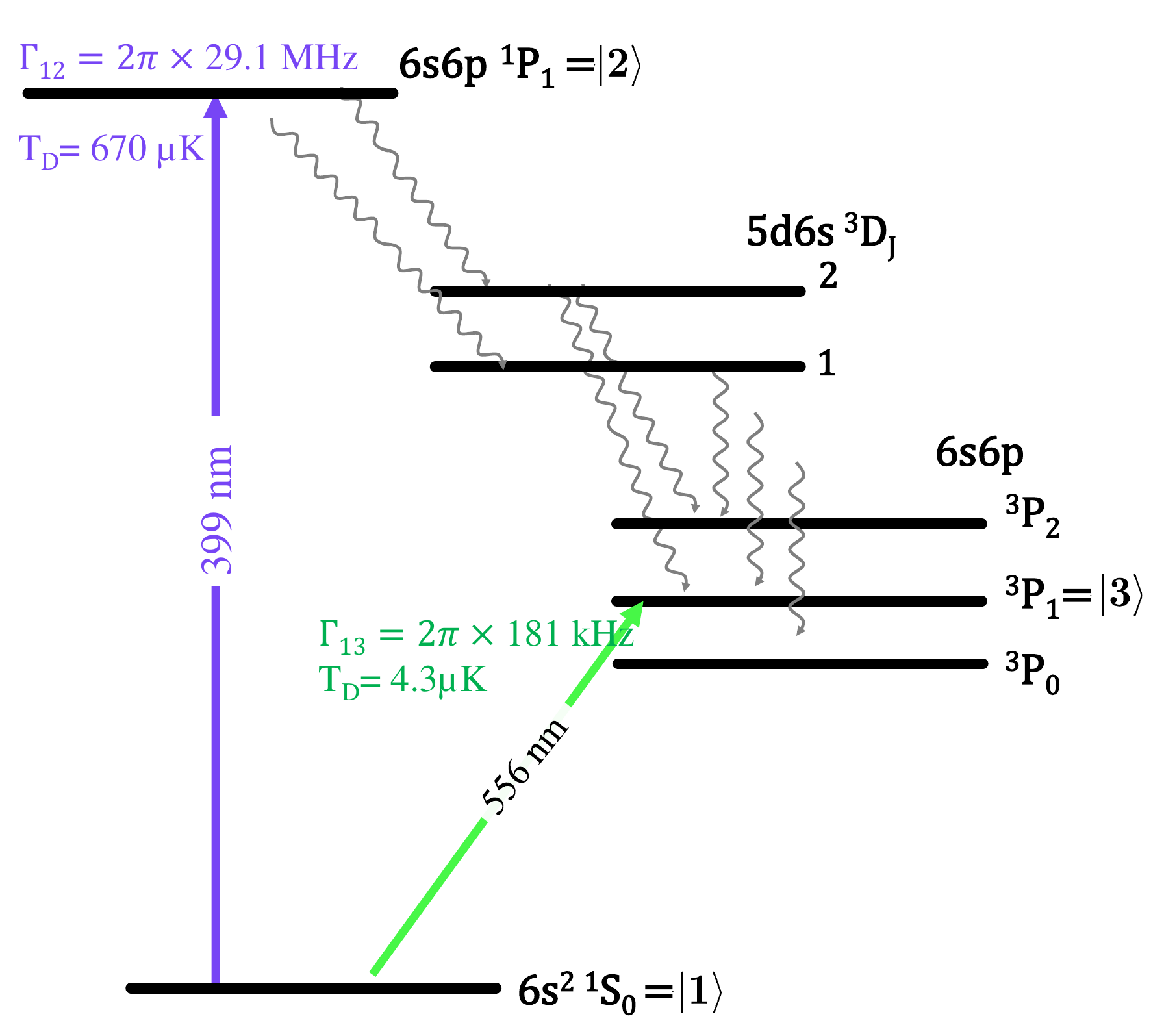}
 	  \caption{\label{EScheme}\justifying The relevant energy levels of Yb. $\Gamma$  and T$_{\textrm{D}}$ are the linewidth and the Doppler temperature of the transition. The wavy arrow shows the decay paths of the $^1$P$_1$ state.}
 \end{figure}

\subsection{Laser system and optics scheme}

\begin{figure}[htb]
 \centering
 \includegraphics[width=\linewidth]{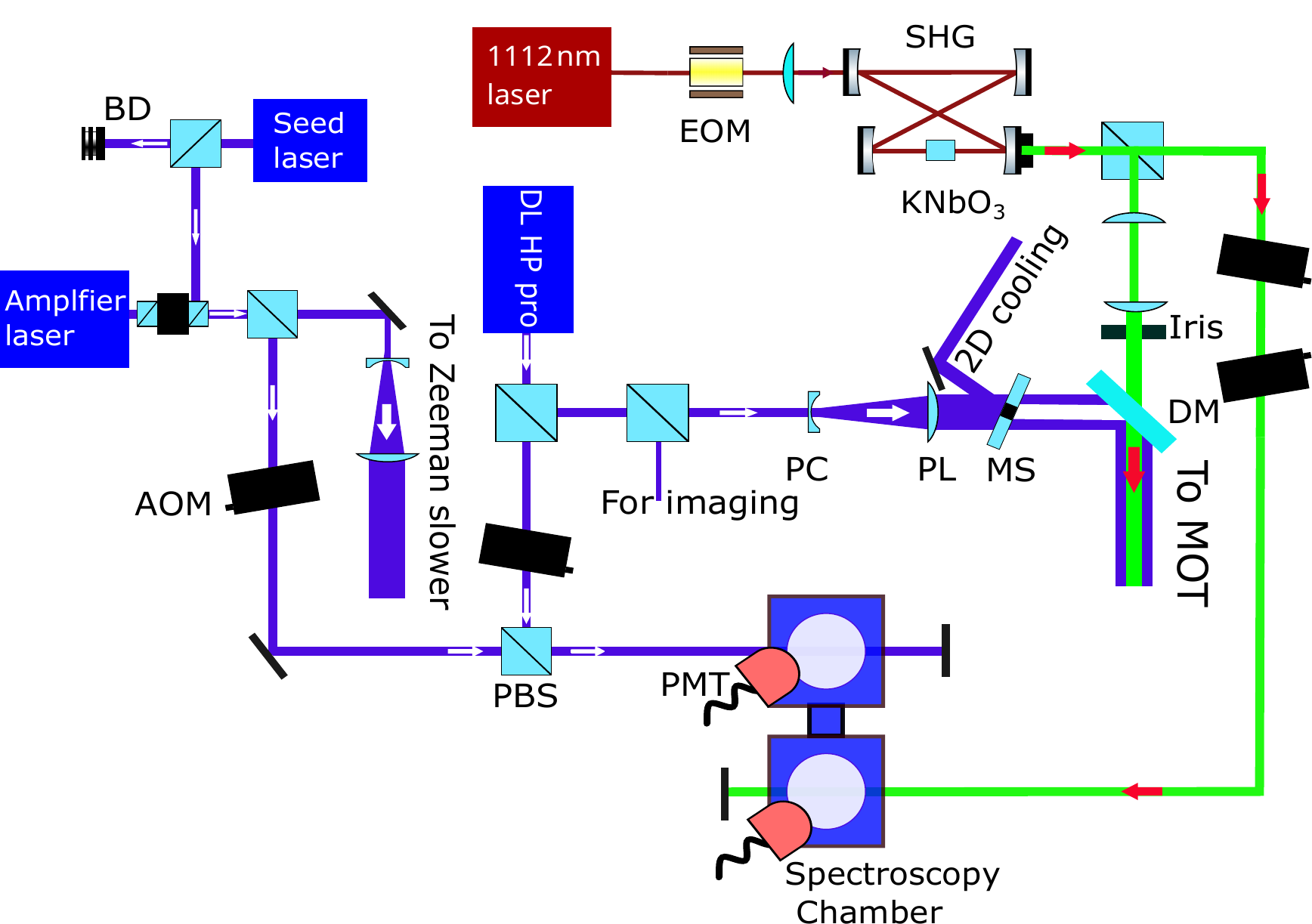}

 \caption{\justifying \label{lscheme}Optics layout for laser system. Figure abbreviations: PL$-$ Plano-convex lens, PC$-$Plano-concave lens, DM$-$ Dichroic mirror, PMT$-$ Photomultiplier tube, BD$-$ Beam dump, MS$-$Mask, PBS$-$Polarising beam splitter}.
 \end{figure}

The relevant energy levels scheme of Yb is shown in Fig. \ref{EScheme}. It possesses two cooling transitions: a strong, broad singlet transition at 399 nm, 6s$^2$ $^1$S$_0$ $\rightarrow$ 6s6p $^1$P$_1$ and a weak, narrow-line triplet transition at 556 nm, 6s$^2$ $^1$S$_0$ $\rightarrow$ 6s6p $^3$P$_1$ . The optics layout for the laser systems is shown in Fig. \ref{lscheme}.
We use two separate laser systems at 399 nm for blue MOT and Zeeman slower. The blue MOT laser system is a commercially available ECDL (Make: Toptica, Model: DL Pro HP) with a maximum output power of 130 mW. The laser beam is divided into two portions, one for spectroscopy and another for the MOT setup. The spectroscopy portion is up-shifted in frequency by 56 MHz using an AOM and sent to the spectroscopy chamber at 90$^{o}$  with respect to the atomic beam direction to remove the Doppler broadening. The Zeeman slower laser is a commercially available laser system (Make: MogLabs, Model: ILA) that operates based on active injection locking mechanism \cite{gupta_2016}. It comprises two lasers, a low-power stable seed laser and a high-power amplifier diode laser. A small portion of the seed beam is injected into the amplifier diode laser to get a high power output with similar spectral features as the seed laser. The Zeeman slower beam, which counter-propagates to the atomic beam, has a beam diameter of 16 mm at the entrance window and converges to 6 mm at the nozzle. It has power of 65 mW and is red detuned by 500 MHz.

\begin{figure}
\centering
\includegraphics[width=\linewidth]{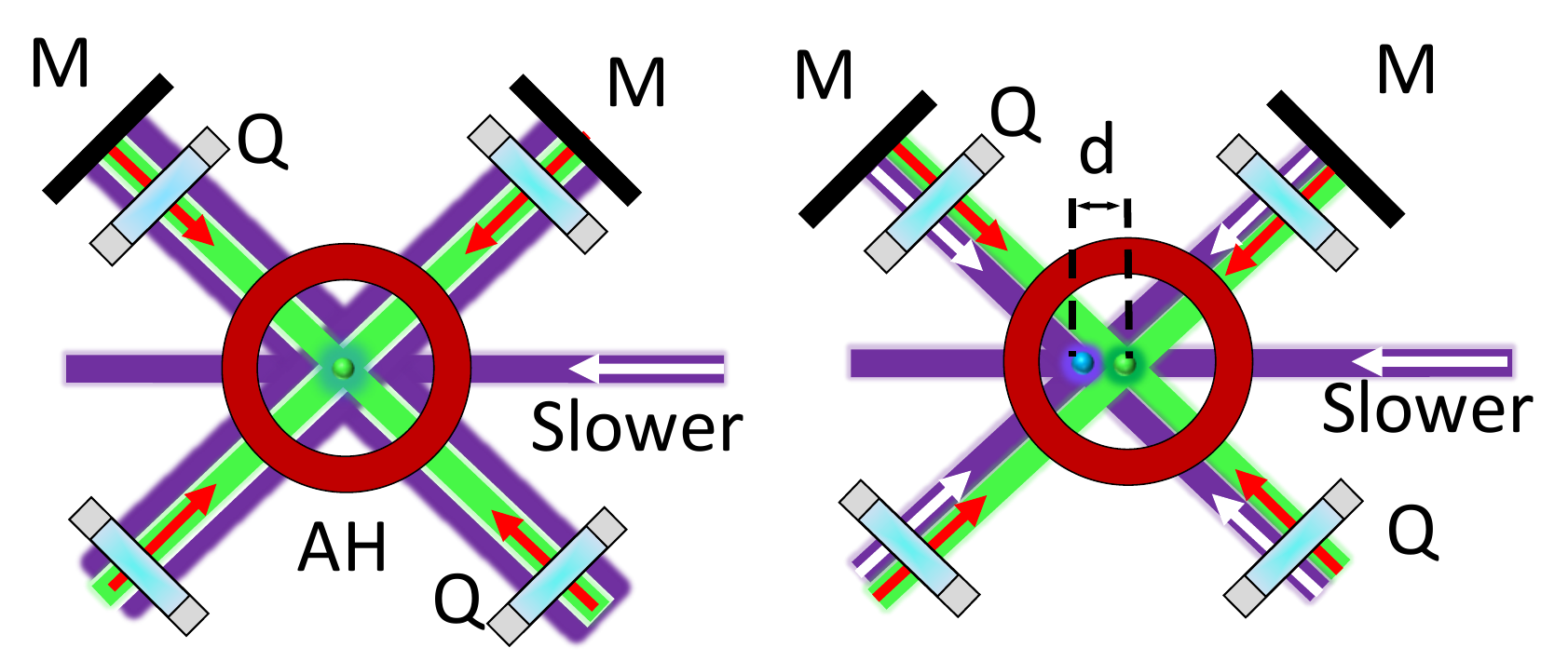}
	\begin{picture}(0,0)
		\put(-120,112){(a)}
		\put(00,112){(b)}
	\end{picture}
\includegraphics[width=\linewidth]{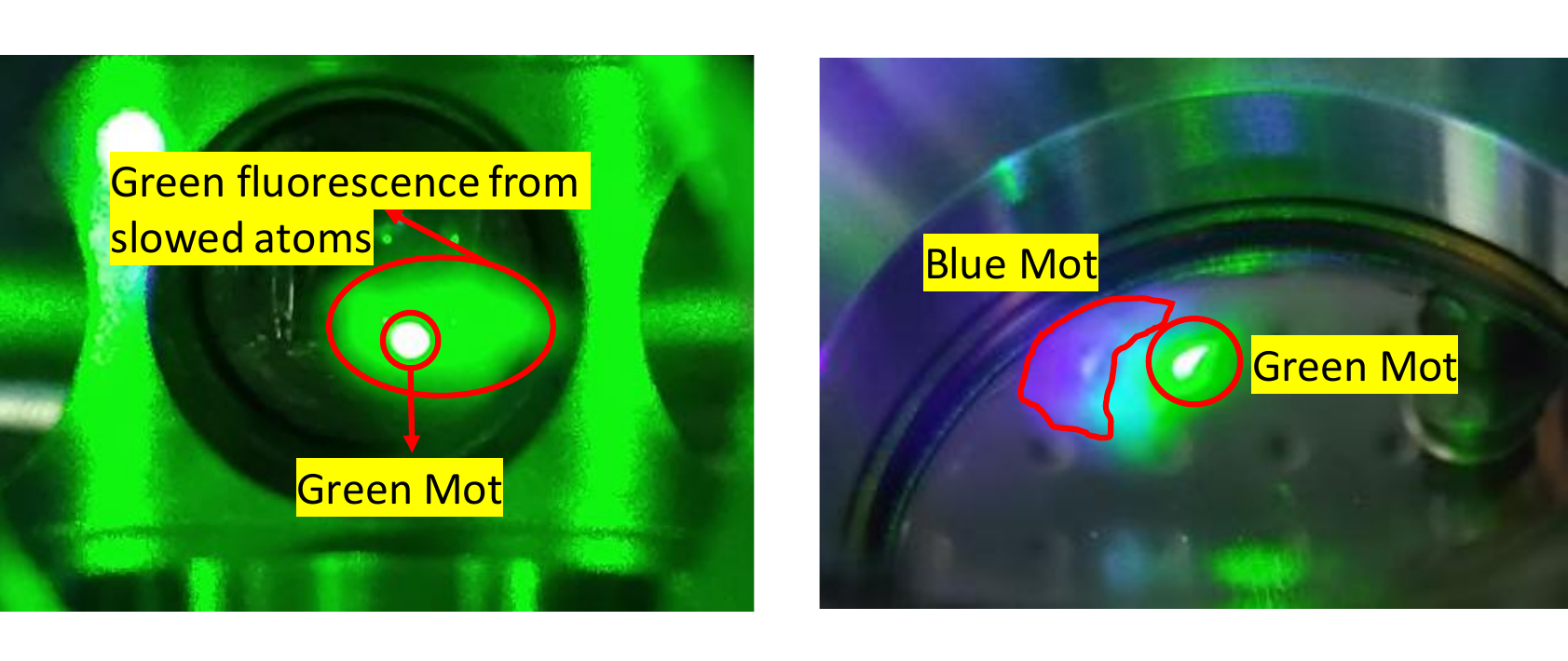}
	\begin{picture}(0,0)
		\put(-120,110){(c)}
		\put(00,110){(d)}
	\end{picture}
\caption{\label{Mscheme} \justifying Top view of the MOT set-up of  (a) core-shell and (b) center-shifted dual MOT configuration. Figure abbreviations: AH $-$ Anti-Helmholtz coil, M$-$ Mirror, Q $-$ Dual  $\lambda/4$ wave-plate, $d$ $-$ distance between blue and green MOT. The 399 nm and 556 nm beams are shown in violet and Green colour, respectively. Images of green MOT in core-shell configuration (c) and center-shifted dual MOT configuration (d)}.
\end{figure}

The 556 nm laser is generated (typically 10 mW) by frequency doubling using a  Potassium Niobate (KNbO$_{3}$) crystal inside a bow-tie cavity. The fundamental laser at 1111~nm with a typical power of 200~mW is generated using an ECDL made of a diode laser (Make: Innolume, Model: SM-1120-TO-500). The 556 nm laser beam is further divided into two portions using PBS for spectroscopy and MOT. A pair of AOMs is used on the path of the spectroscopy beam for the MOT detuning and the frequency modulation required for the laser stabilization.

The MOT mixing scheme for core-shell configuration is shown in Fig. \ref{Mscheme}(a). A hole is created at the center of the blue beam by passing it through a circular mirror attached to a glass window, which is placed in its path. It reflects the central portion of the beam and utilized for 2D cooling. This approach allows the unused (masked) power to be recycled for experiments. The hole is then filled with green light of a desired size using a dichroic mirror. An iris is placed in the path of the green beam to control its beam size. The mixed beams are further split into three beams using PBSs and sent to the MOT chamber. Mechanical shutters are used in the paths of lasers for switching the lasers off and on. 

In the center-shifted dual MOT configuration, the blue beam and green beam are aligned side-by-side as shown in Fig. \ref{Mscheme}(b). The blue and green beams have the same beam diameter of 9 mm with powers of 30 mW and 9 mW, respectively. The blue beam diameter is limited to 9 mm due to constraints of the optical access of the MOT chamber windows. The overlapping region of green MOT beams is aligned to coincide with the region of zero magnetic field, while the blue beams create the blue MOT between the green MOT and the exit of the Zeeman slower, positioned closer to the green MOT at a non-zero magnetic field. 

 We use absorption imaging analysis with a CMOS camera (Make: Thorlabs, Model: CS135MUN) to measure the number of atoms in the Green MOT.

\section{Theory} \label{Theory}

We theoretically study the capture dynamics of Yb atoms in both the core-shell and center-shifted dual MOT configurations using density matrix formalism in one dimension. 
In our model, we consider a simplified three-level atomic system, where the ground state $^1S_0 = \ket{1}$ is coupled to two excited states: $^1P_1 = \ket{2}$ via the 399~nm transition, and $^3P_1 =\ket{3}$ via the 556~nm inter-combination line. $^1S_0 \rightarrow ^3P_1$ transition is a close transition while $^1S_0 \rightarrow ^1P_1$ transition is open. The state $^1P_1$ decays out of the transition cycle at rate $2\pi \times$ 1.03 Hz (6.48 s$^{-1}$) \cite{YbOpPump}.

The atom-light interaction Hamiltonian for laser beams propagating in the $\pm z$ direction is given by,
\begin{align}
H_{\pm} = \, & \hbar\Big[-(\Delta_{12} \mp k_{12} v_z \mp \mu_B B(z))\ket{2}\bra{2} \nonumber \\
& -(\Delta_{13} \mp k_{13} v_z \mp \mu_B B(z)\ket{3}\bra{3} \nonumber \\
& + \frac{\Omega_{12}}{2}\ket{1}\bra{2} + \frac{\Omega_{13}}{2}\ket{1}\bra{3} + \text{h.c.} \Big]
\label{eqHa}
\end{align}
where $\Delta_{12}$ and $\Delta_{13}$ are the laser detunings, $k_{12} = 2\pi / 399~\mathrm{nm^{-1}}$ and $k_{13} = 2\pi  / 556~\mathrm{nm^{-1}}$ are the wavevectors, $\Omega_{12}$ and $\Omega_{13}$ are the Rabi frequencies, and $\mu_B B(z)$ represents the Zeeman shift due to the magnetic field gradient.

To understand the system dynamics, we solve the Liouville--von Neumann master equation,
\begin{align}
\frac{d\rho_\pm}{dt} = -\frac{i}{\hbar}[H_\pm, \rho_\pm] + L(\rho_\pm)
\label{eqlv}
\end{align}
where $\rho$ is the density matrix and $L(\rho)$ denotes the Lindblad operator that accounts for spontaneous decay.

Solving Eq.~\ref{eqlv} yields the time-dependent density matrix elements $\rho_{ij}$, from which we extract observables such as absorption and scattering forces. The imaginary parts of the coherence $\rho_{12}^{\pm}$ and $\rho_{13}^{\pm}$ give absorption of the 399~nm and 556~nm laser beams propagating in the $\pm z$ directions, respectively.

The net radiation force experienced by the atom is expressed as,
\begin{align}
\label{eqForce}
m\frac{dv_z}{dt} = \hbar \Big[ k_{12} \Omega_{12} \, \mathrm{Im}(\rho_{12}^+ - \rho_{12}^-) \\ \nonumber
+ k_{13} \Omega_{13} \, \mathrm{Im}(\rho_{13}^+ - \rho_{13}^-) \Big] 
\end{align}
\begin{figure}
    \centering
    \includegraphics[width=\linewidth]{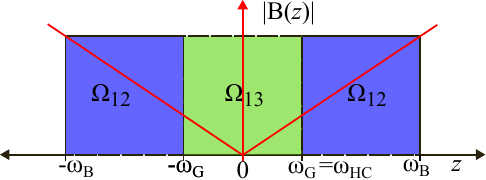}

    \includegraphics[width=\linewidth]{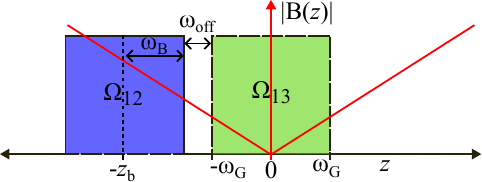}
     \begin{picture}(0,0)
		\put(-100,190){(a)}
        \put(-100,95){(b)}
	\end{picture}

\caption{  \label{figspb} \justifying Illustration of spatial boundary for (a) core-shell and (b) center-shifted MOT configurations. The parameters for core-shell are, $w_{\mathrm{B}} =9~\mathrm{mm}~( \textrm{half width of the blue beam})$, $w_{\mathrm{HC}}=3~\mathrm{mm} $ (half width of the hollow core) and $w_{\mathrm{G}}=3~\mathrm{mm}$ (half width of the green beam) and for center-shifted are, $w_{\mathrm{B}} =9~\mathrm{mm}~( \textrm{half width of the blue beam})$, $w_{\mathrm{off}}=1~\mathrm{mm} $ (gap between green and blue beam) and $w_{\mathrm{G}}=3~\mathrm{mm}$ (half width of the green beam) and $ z_b = w_{\mathrm{G}} + w_{\mathrm{off}} + w_{\mathrm{B}}$.}
   
\end{figure}

\begin{figure}[t]
	\includegraphics[width=0.6\linewidth]{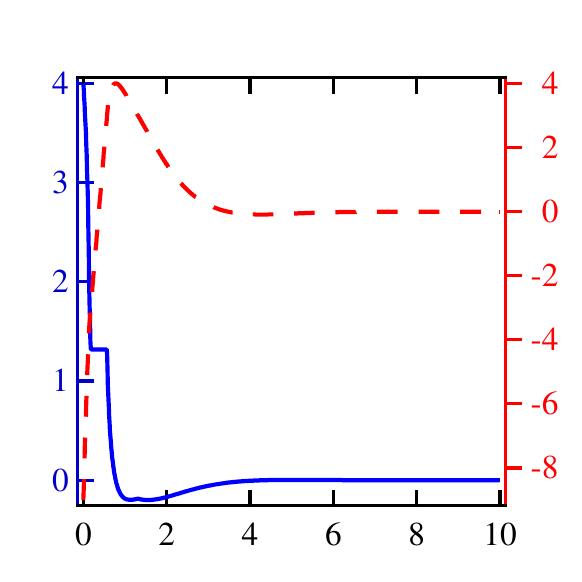}
	\begin{picture}(1,0)
		\put(-170,130){(a)}
		\put(-160,60){\rotatebox{90}{$k_{12} v_z/\Gamma_{21}$}}
		\put(-85,0){t (ms)}
        \put(5,60){\rotatebox{90}{z (mm)}}
	\end{picture}
 	\includegraphics[width=0.6\linewidth]{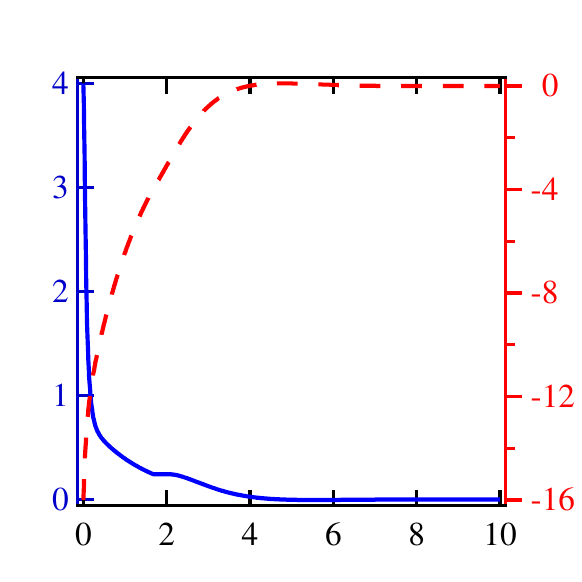}
	\begin{picture}(1,0)
		\put(-170,130){(b)}
		\put(-160,60){\rotatebox{90}{$k_{12} v_z/\Gamma_{21}$}}
		\put(-85,0){t (ms)}
        \put(5,60){\rotatebox{90}{z (mm)}}
	\end{picture}
 
    \caption{ \label{figvpt} \justifying Velocity (blue solid line) and position (red dashed line) trajectory with time for (a) core-shell and (b) center-shifted configurations for $v_{in}$$=4\Gamma_{21}/k_{12}$.  The simulation parameters for core-shell are, $\Omega_{13} = 10 \Gamma_{13}$, $\Omega_{12}=1\Gamma_{12}$, $B = 12$ G/cm, $\Delta_{13}= -10\Gamma_{13}$, $\Delta_{12}=-2\Gamma_{12}$ and for center-shifted are, $\Omega_{13} = 10\Gamma_{13}$, $\Omega_{12}=1.5\Gamma_{12}$, $B = 12$ G/cm, $\Delta_{13}= -10\Gamma_{13}$, $\Delta_{12}=-2\Gamma_{12}$,  $w_{\mathrm{off}} = 1$ mm.}
\end{figure}

\begin{figure}[t]
\includegraphics[width=0.85\linewidth]{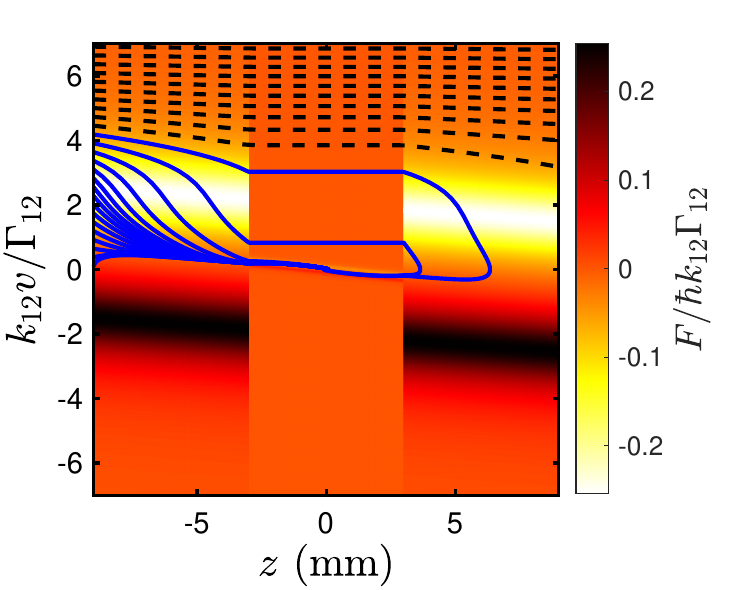}
\put(-130,70){\textcolor{green}{$\times$ 200}}\\
\includegraphics[width=0.85\linewidth]{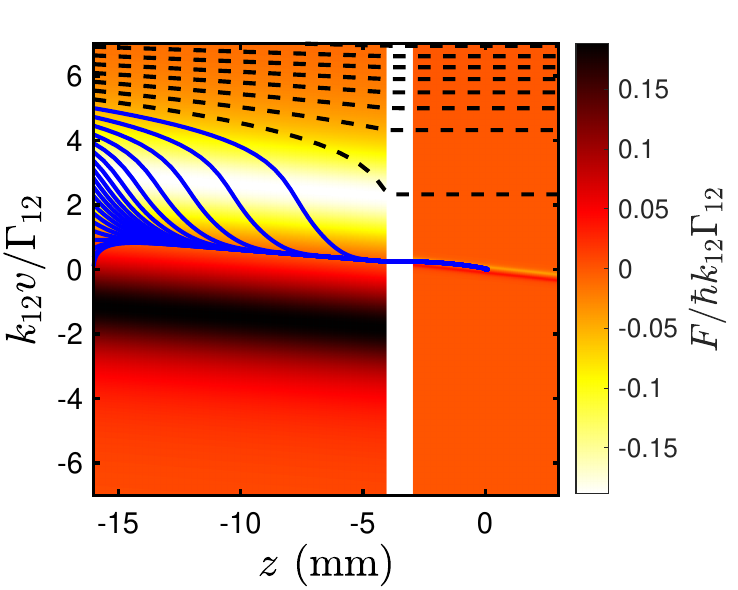}
\begin{picture}(0,0)
 \put(-90,70){\textcolor{green}{$\times$ 200}}
\end{picture}

\caption{ \label{figTraject}\justifying Force vs velocity and position along with atoms phase space trajectory (blue solid (trapped) and black dashed (untrapped) curve) for core-shell (a) and center-shifted (b) configurations. The simulation parameters for core-shell are, $\Omega_{13} = 10\Gamma_{13}$, $\Omega_{12}=1\Gamma_{12}$, B = 12 G/cm, $\Delta_{13}= -10\Gamma_{13}$, $\Delta_{12}=-2\Gamma_{12}$ and for center-shifted are, $\Omega_{13} = 10\Gamma_{13}$, $\Omega_{12}=1.5\Gamma_{12}$, B = 12 G/cm, $\Delta_{13}= -10\Gamma_{13}$, $\Delta_{12}=-2\Gamma_{12}$,  $w_{\mathrm{off}} =$ 1 mm.} 
\end{figure}

We simulate the velocity and position trajectories of atoms in both the core-shell and center-shifted MOT configurations by defining spatially dependent boundaries for the laser fields as shown in Fig. \ref{figspb}.


In the core-shell configuration, the 399~nm (blue) beam is assumed to be effective only in the outer shell region, while the 556~nm (green) beam is confined to the central core region. The magnitude of the magnetic field is zero at $z=0$. This is implemented in the simulation by defining the position-dependent Rabi frequencies $\Omega_{12}(z)$ and $\Omega_{13}(z)$ as a piecewise:
\begin{align}
\Omega_{12}(z) &=
\begin{cases}
\Omega_{12}, & \text{if }  w_{\mathrm{B}} > |z| > w_{\mathrm{HC}} \\
0, & \text{otherwise}
\end{cases} \\
\Omega_{13}(z) &=
\begin{cases}
\Omega_{13}, & \text{if } |z| \le w_{\mathrm{G}} \\
0, & \text{otherwise}
\end{cases}
\end{align}

Here, $w_{\mathrm{B}}=9~\mathrm{mm}$, $w_{\mathrm{HC}}=3~\mathrm{mm}$ and $w_{\mathrm{G}}=3~\mathrm{mm}$  corresponds to the half width of the blue beam, the hollow core on the blue beam and the green beam, respectively as shown in Fig. \ref{figspb}(a). This piecewise definition captures the essential structure of the core-shell MOT, where blue beam surrounds a central void filled with green light. 

In the center-shifted MOT configuration, the spatial profiles of the laser beams are modeled as rectangular windows as shown in Fig. \ref{figspb}(b). The green MOT beam has a full width of \( 2w_{\mathrm{G}} ~(=6~\mathrm{mm})\) and is centered at \( z = 0 \). The blue MOT beam has a full width of \( 2w_{\mathrm{B}} ~(=12~\mathrm{mm})\) and is located entirely to the left of the green MOT. There is a spatial gap of width \( w_{\mathrm{off}} = 1~ \mathrm{mm}\) introduced between the two beams. The blue beam full width is taken to be 12 mm, which corresponds to the effective blue beam width in the core-shell configuration after introducing a 6 mm hollow core.
Here, the position-dependent Rabi frequencies are defined as:
\begin{align}
\Omega_{12}(z) &=
\begin{cases}
\Omega_{12}, & \text{if } z_b - w_{\mathrm{B}} < z < z_b + w_{\mathrm{B}} \\
0, & \text{otherwise}
\end{cases}\\
\Omega_{13}(z) &=
\begin{cases}
\Omega_{13}, & \text{if } -w_{\mathrm{G}} < z < w_{\mathrm{G}} \\
0, & \text{otherwise}
\end{cases}
\end{align}
\(\\\mathrm{where},~ z_b = -(w_{\mathrm{G}} + w_{\mathrm{off}} + w_{\mathrm{B}}).\)
With these spatial conditions, we numerically solve the coupled equations derived from Eqs. \ref{eqlv} and \ref{eqForce} to track the evolution of atomic position $z(t)$ and velocity $v_z(t)$ over time. For all simulation plots, $\Omega_{12}$ for the center-shifted configuration is set to 1.5 times that of the core-shell configuration due to the ratio of the respective blue beam widths.\\

The simulated trajectory for the velocity and position vs time for the atom with initial velocity $v_{in}$$=4\Gamma_{21}/k_{12}$ are shown in Fig. \ref{figvpt}(a) for core-shell configuration and in Fig.~\ref{figvpt}(b) for center-shifted MOT configuration. In both cases, atoms experience rapid deceleration within the first 5 ms due to the strong damping force exerted by the blue beam. Because the slowing happens so quickly, the chance of atoms being lost to metastable states is very less. After this initial slowing, the atoms enter the green MOT region, where further cooling and trapping occur via the narrow 556 nm transition.

The atomic phase space trajectories (blue solid and black dashed curves) for the various initial velocities and the corresponding force profile (2D color plot) are shown in Fig.~\ref{figTraject}(a) for core-shell configuration and Fig.~\ref{figTraject}(b) for center-shifted MOT configuration. The blue curve represents the trajectory of an atom that is successfully captured and trapped within the MOT, while the black curve corresponds to an untrapped atom that escapes the MOT region. The condition for captured atom is, final position should be within the green region and final velocity should be below Doppler cooling limit of the green transition. The simulation was run for total duration of t$ =1 ~\textrm{s} $ with time step of $\delta$t $=$$10^{-5} ~\textrm{s}$.  The force exerted by the green MOT beam is significantly weaker compared to that of the blue beam due to its narrower linewidth. To enhance its visibility in the plot, the force profile in the green region is artificially scaled by a factor of 200. 

\begin{figure}[t]
\includegraphics[width=0.8\linewidth]{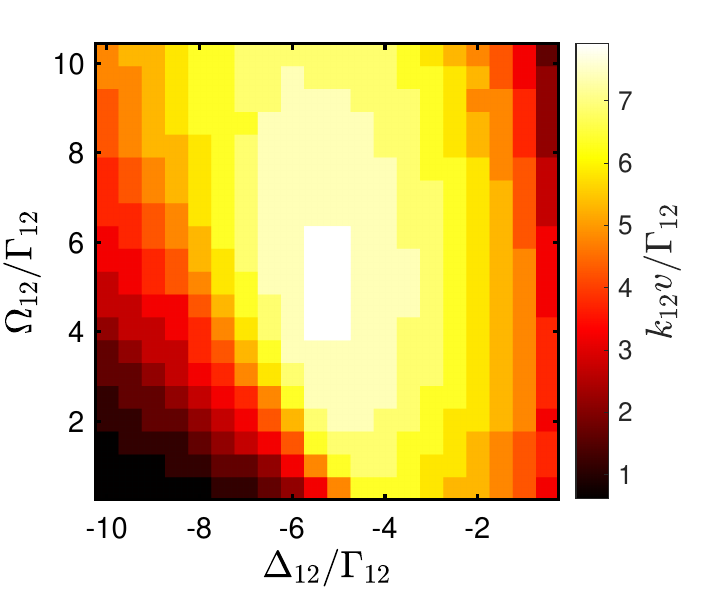}
\includegraphics[width=0.8\linewidth]{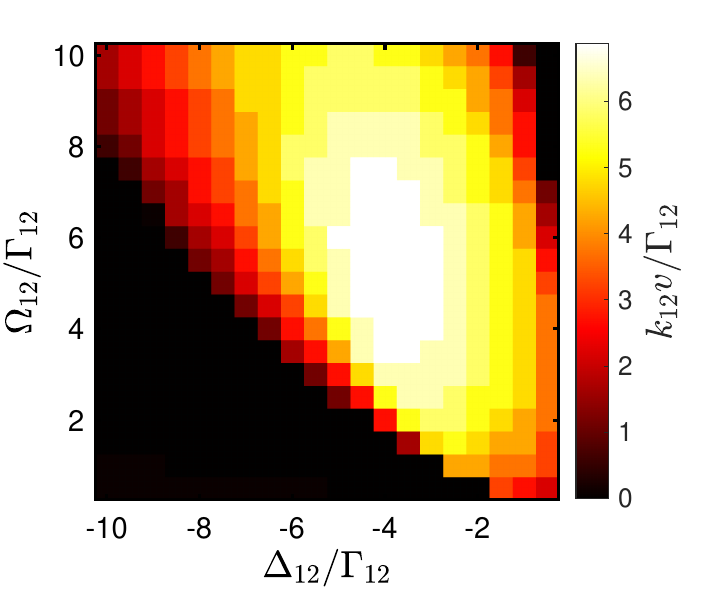}

\caption{ \label{figCapv} \justifying Capture velocities vs detuning ($\Delta_{12}$) and Rabi frequency ($\Omega_{12}$) of blue laser for (a) core-shell and (b)  center-shifted configurations. The simulation parameters for core-shell are, $\Omega_{13} = 10 \Gamma_{13}$,   B = 12 G/cm, $\Delta_{13}= -10\Gamma_{13}$. and for center-shifted are, $\Omega_{13} = 10\Gamma_{13}$,  B = 12 G/cm, $\Delta_{13}= -10\Gamma_{13}$, $w_{\mathrm{G}} =$ 12 mm, $w_{\mathrm{B}} =$ 6 mm, $w_{\mathrm{off}} =$ 1 mm.}
\end{figure}

The simulated capture efficiency vs detuning and Rabi frequency of blue laser is shown in Fig. \ref{figCapv}(a) for core-shell configuration and in Fig.~\ref{figCapv}(b) for center-shifted MOT configuration. The maximum capture velocity captured in the core-shell and center-shifted MOT configuration are $\approx$ 8$\Gamma_{12}/k_{12}$ and $\approx $ 7$\Gamma_{12}/k_{12}$, respectively. 

\begin{figure}[t]
	\includegraphics[width=0.6\linewidth]{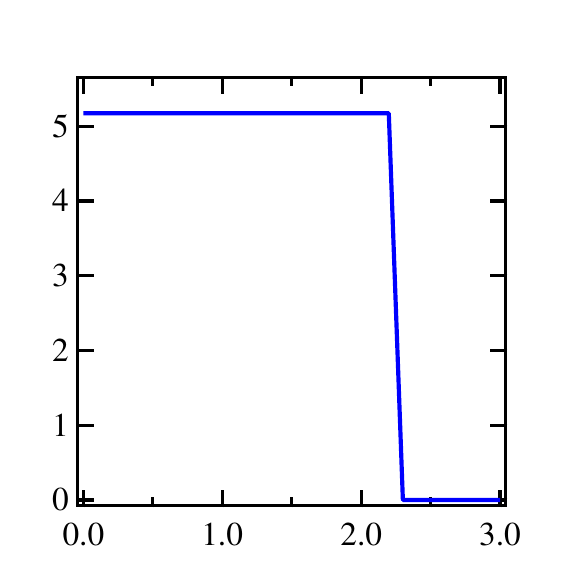}
	\begin{picture}(1,0)
		\put(-170,130){(a)}
		\put(-160,60){\rotatebox{90}{$k_{12} v_z/\Gamma_{21}$}}
		\put(-95,0){$w_{\mathrm{off}}$ (mm)}

	\end{picture}

    \caption{ \label{figvvssep} \justifying Capture velocity vs separation between the blue and green beam in center-shifted MOT configuration. The simulation parameters are,  $\Omega_{13} = 10\Gamma_{13}$, $\Omega_{12}=2\Gamma_{12}$, $B = 12$ G/cm, $\Delta_{13}= -10\Gamma_{13}$, $\Delta_{12}=-2\Gamma_{12}$.}
\end{figure}

In the experiment, we observed that a larger number of atoms were loaded into the green MOT in the core-shell configuration as compared to the center-shifted one, although the intensity of blue beam is higher in the latter configuration. This is because the center-shifted MOT is very sensitive to the separation between the blue beam and the green beam region, as shown in Fig. \ref{figvvssep}. It is to be noted that when the separation exceeds 2.2 mm, even atoms with zero initial velocity are not captured. This is because the magnetic field gradient accelerates the atoms toward the center, which stops them from being captured by the green MOT.

\section{Results and Discussion}{\label{RnD}}
The table \ref{table} summarizes the number of atoms loaded in green MOT with various configurations. The blue beam with a diameter, $\phi_\mathrm{B}=$ 18 mm and the hollow core (HC) diameter, $\phi_{\textrm{HC}} =$ 6 mm gives the maximum number of atoms. The total power in the blue MOT beam after creating a hollow core is 30 mW. We reflect 12 mW of blue power that can be used for 2D cooling. We now discuss our experimental results of the green MOT in the core-shell configuration.

\begin{figure}[t]
	\centering
	\includegraphics[width=0.45\linewidth]{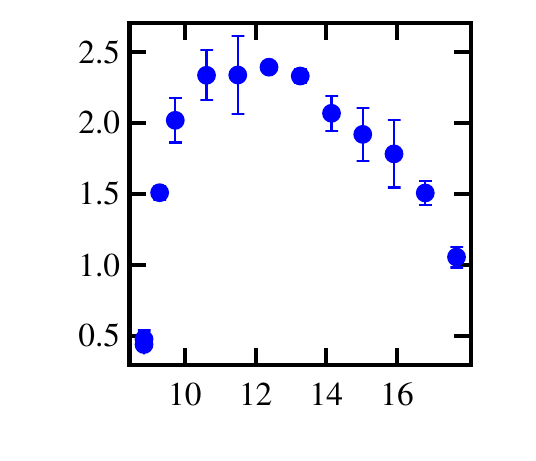}
	\hfill
	\includegraphics[width=0.45\linewidth]{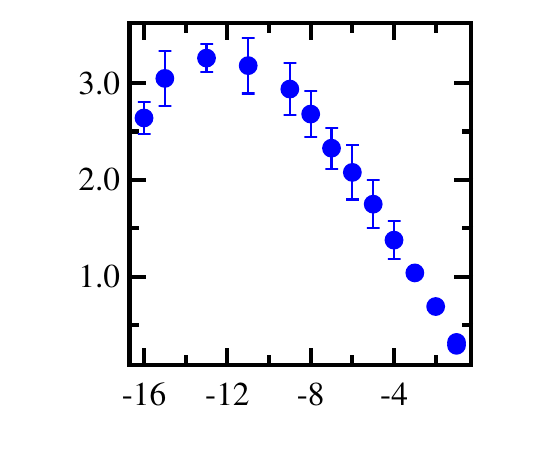}	
		\begin{picture}(0,0)
		\put(-120,85){(b)}
		\put(-120,45){\rotatebox{90}{$N/10^{8}$}}
		\put(-78,-5){$\Delta_{13}/\Gamma_{13}$ }
		\put(-250,85){(a)}
		\put(-250,45){\rotatebox{90}{$N/10^{8}$}}
		\put(-210,-5){$B' \textrm{(G/cm)}$}
	\end{picture}
	
    	\caption{\label{NvsID}\justifying  Number of atoms ($N$) in the green MOT vs (a) magnetic field gradient ($B'$)  and (b) detuning ($\Delta_{13}$) of the green laser. Experimental parameters: blue MOT laser power, $P_{399}=25$ mW, green MOT laser power, $P_{\textrm{556}}=7$ mW, blue MOT laser detuning, $\Delta_{12}=-2 \Gamma_{12} $, blue  beam diameter, $\phi_{\mathrm{B}}= 18$ mm and the diameters of the hollow core and the green beam are both $\phi_{\mathrm{G/HC}}= $ 6 mm. }
\end{figure}
\begin{figure}[t]
	\centering
    \includegraphics[width=0.45\linewidth]{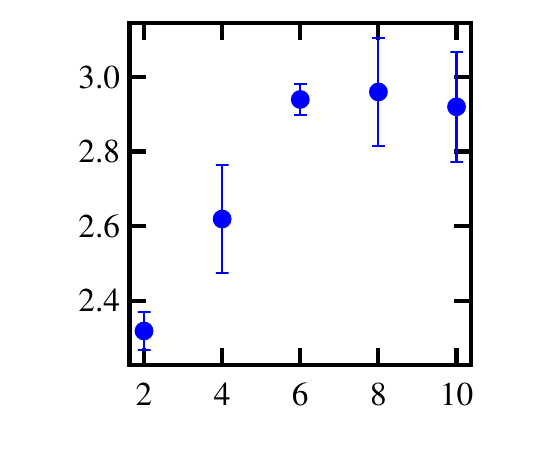}
    \begin{picture}(0,0)
		\put(-120,45) {\rotatebox{90}{N/$10^{8}$}}
		\put(-70,0){$\phi_{\textrm{G}}$ (mm)}
	\end{picture}
    \includegraphics[width=0.45\linewidth]{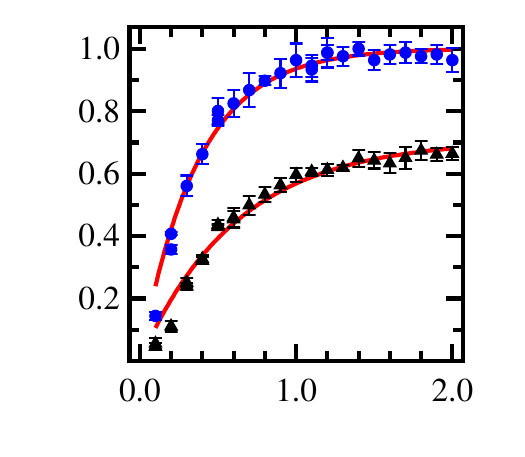}
        \begin{picture}(0,0)
        \put(-250,85){(a)}
        \put(-120,85){(b)}
		\put(-120,45) {\rotatebox{90}{$ N_{\textrm{norm}} $}}
		\put(-70,0){t (s)}
	\end{picture}
	\caption{\label{NvsS}\justifying  (a) Number of atoms ($N$) in the green MOT vs green beam size ($\phi_{\textrm{G}}$) with $\phi_{\mathrm{B}}= 18$ mm, $\phi_{\textrm{HC}}= 6$ mm, and $\Delta_{13}=-10\Gamma_{13}$. (b) Loading curve with $\phi_{\mathrm{B}}= 18$ mm and $\phi_{\textrm{HC/G}}= 6$ mm (blue circle) and 10 mm (blue triangle) with  $\Delta_{13}=-10\Gamma_{13}$. The red solid lines are the corresponding exponential fit.}
\end{figure}

First, we discuss the atom number ($N$) of the green MOT with magnetic field gradient ($B'$) and the green MOT laser detuning ($\Delta_{13}$) as shown in Fig. \ref{NvsID}. $N$ reaches maximum at around $B'=12$~G/cm and then it decreases with further increase in  $B'$ as shown in Fig. \ref{NvsID}(a). For green MOT laser detuning, $N$ is maximum at $\Delta_{13}= -12\Gamma_{13}$ as shown in Fig. \ref{NvsID}(b). 
\begin{table}[t]
    \centering
       \caption{\justifying {No of atoms ($N$) in the green MOT with various configurations.}}
    \begin{tabular}{c c c c }
        \hline
        &$\phi_{\mathrm{B}}$ (mm) &  $\phi_{\textrm{G}} ~\&~\phi_{\textrm{HC}} $ (mm) & $N$ \\
        \hline
        Core-shell MOT &18  & 6   &  3.4 $\times $10$^{8}$ \\
        &18   & 10   & 1.0 $\times $10$^{8}$   \\
        &30   & 16   & 4.5 $\times $10$^{7}$    \\
        \hline
        center-shifted MOT&9&9&1.9 $\times $10$^{7}$ \\
            \hline
    \end{tabular}
   
    \label{table}
\end{table}
Now, we study the variation of the atom number in the green MOT with various green beam sizes in Fig. \ref{NvsS}(a). With an increase in the diameter of the green laser beam ($\phi_\textrm{G}$), the atom number increases and saturates. The saturation point is $\phi_\textrm{G}\approx\phi_\textrm{HC}$. This is because in the presence of both blue and green lasers, the cooling is mainly determined by the blue laser as the blue transition is a stronger transition by two orders of magnitude. Unlike the previous study in Rb, where the weak transition is only 4-5 times weaker, the presence of the weaker transition helps, and the core was overfilled by the weak transition laser  \cite{RCD_CMOT}. For high power of green laser, such that the power broadened linewidth of weak transition is comparable to the broad transition, we may see the effect of overfilling due to cooling by the green transition.

Further, we study the loading rate of the green MOT for the hollow core and green beam sizes $\phi_{\textrm{HC}}~\&~\phi_{\textrm{G}}=$ 6~mm and 10 mm with $\phi_{\mathrm{B}}=$ 18~mm. The loading rate is determined by turning on the blue and green MOT for a given time (loading time) and determining the number of atoms. The loaded atoms, $N$$_\textrm{norm}$ (normalized with maximum number of atoms) vs loading time ($t$) is shown in Fig. \ref{NvsS}(b) by blue solid circle ($\phi_{\textrm{HC}}~\&~\phi_{\textrm{G}}=$ 6~mm) and triangular points ($\phi_{\textrm{HC}}~\&~\phi_{\textrm{G}}=$ 10~mm). The curve is fitted with $N(t)=N_0(1-e^{-t/t_\textrm{L}})$ and shown in red curve. The loading time ($t_\textrm{L}$) for these two configurations, $\phi_{\textrm{HC}}~\&~\phi_{\textrm{G}}=$ 6~mm and $\phi_{\textrm{HC}}~\&~\phi_{\textrm{G}}=$ 10~mm with $\phi_\mathrm{B}=$18~mm, are 0.4 s and 0.6 s respectively.

\begin{figure}[t]	
	\centering
	\includegraphics[width=0.45\linewidth]{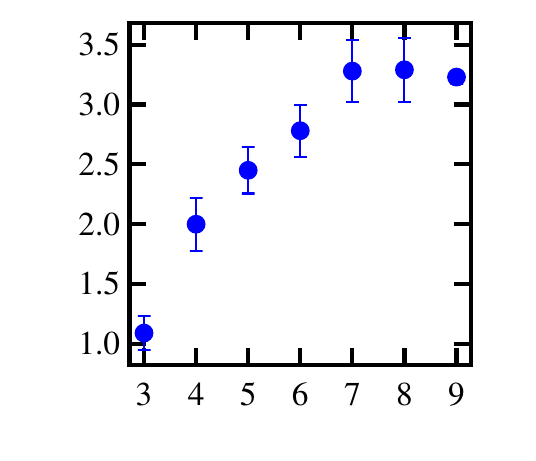}
	\hfill
	\includegraphics[width=0.45\linewidth]{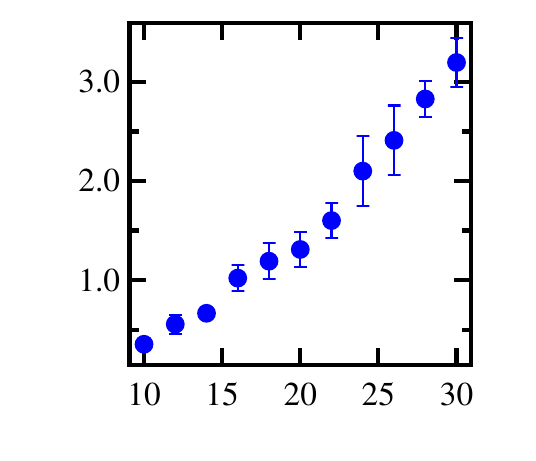}	
	\begin{picture}(0,0)
		\put(-120,85){(b)}
		\put(-120,45){\rotatebox{90}{$N/10^{8}$}}
		\put(-70,-5){$P_{399}$ (mW)}
		\put(-250,85){(a)}
		\put(-250,45){\rotatebox{90}{$N/10^{8}$}}
		\put(-210,-5){$P_{556}$ (mW)}
	\end{picture}
	\caption{\label{NvsP}\justifying (a) Number of atoms ($N$) in the green MOT vs (a) Green beam power ($P_{399}$) and (b) Blue beam power ($P_{556}$). Experimental parameters: $\Delta_{12}=-2 \Gamma_{12} $ , $\Delta_{13}=-10\Gamma_{13}$, $\phi_{\mathrm{B}}= 18$ mm and $\phi_{\textrm{HC}}~\&~\phi_{\textrm{G}}=$ 6 mm.}
\end{figure}

We further study $N$ with the power of the green MOT and blue MOT lasers, as shown in Fig. \ref{NvsP}. $N$ versus green MOT laser power ($P_{\textrm{556}}$) shown in Fig. \ref{NvsP}(b) shows saturation, suggesting that the available green laser power is sufficient for trapping.
In contrast, the $N$ versus blue laser power ($P_{\textrm{399}}$) curve in Fig. \ref{NvsP}(a) shows a linear relationship, indicating that increasing the blue MOT laser power continues to increase the number of atoms, indicating higher power is desirable. 

\begin{figure}[h!]
	\centering
	\includegraphics[width=0.45\linewidth]{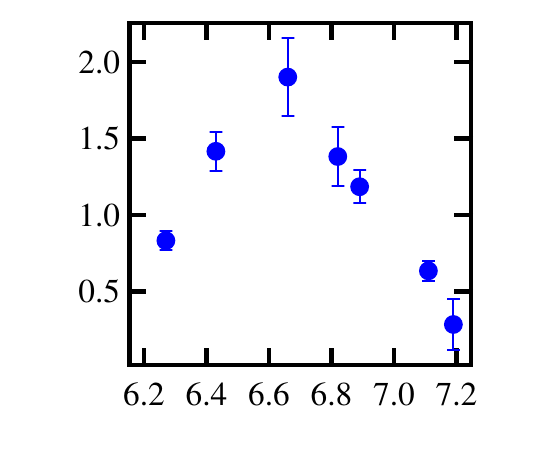}
	\begin{picture}(0,0)
		\put(-120,45) {\rotatebox{90}{N/$10^{7}$}}
		\put(-70,0){$d$ (mm)}
	\end{picture}
	\caption{\label{cshift}\justifying Number of atoms ($N$) in the green MOT vs separation ($d$) between green and center-shifted blue MOT for the optimum experimental parameters.}
\end{figure}
In the core-shell configuration, the central core of the blue beam is masked (even though it is redirected for use in 2D cooling), resulting in a limitation of the available blue power for MOT formation. To overcome this limitation and fully utilize the blue MOT beam power, we implement center-shifted dual MOT configuration. In this configuration, the overlap region of the three counter-propagating blue laser beams is shifted towards Zeeman slower, where the magnetic field is non-zero. The atoms emerging from the Zeeman slower are first pre-cooled and partially trapped in the blue MOT. These atoms subsequently enter the green MOT (zero magnetic field) region and are trapped and cooled further.
However, in this method, the maximum number of atoms we loaded in the green MOT is around 2 $\times$ 10$^{7}$ atoms, which is one order of magnitude lower than the number of atoms trapped in the core-shell MOT configuration. The variation of the number of atoms in the green MOT  with respect to the separation ($d$) between the green MOT and the center-shifted blue MOT is shown in Fig. \ref{cshift}. This separation is controlled by adjusting the MOT center using a shim coil, since the green MOT is more sensitive to the magnetic field than the blue MOT. Even a small displacement significantly affects the atom number in the green MOT.

\section{Conclusion}{\label{Conclusion}}

We have loaded Yb atoms at narrow linewidth green transition using both core-shell and center-shifted dual MOT configurations with limited green laser power of approximately 10 mW. In the core-shell configuration, the green beam is superimposed onto a central hole within the core of the blue beam.  Additionally, the center-shifted MOT configuration was implemented, allowing the entire power of the blue beam to be utilized in the MOT without any loss. However, the core-shell configuration was able to trap a higher number of atoms, with $3.4\times10^{8}$ atoms compared to $2 \times10^{7}$ atoms in the center-shifted MOT configuration. This study will be useful for portable optical clocks using Yb and also for quantum computation using Yb tweezer arrays.

\begin{acknowledgments}
 K.P. would like to acknowledge the funding from DST through Grant No. DST/ICPS/QuST/Theme-3/2019.
\end{acknowledgments}
	
	\bibliography{GreenMOTv1}
	
	\end{document}